# Thermodynamic Properties of Weakly Anisotropic Disordered Magnetic Chains in the Large-S Limit

I. Avgin

Department of Electrical and Electronics Engineering, Ege University,

Bornova 35100, Izmir, Turkey

and

D. L. Huber

Department of Physics, University of Wisconsin-Madison, Madison, WI 53706, USA

## Abstract

We investigate the thermodynamic properties of weakly anisotropic disordered magnetic chains where the nearest-neighbor Heisenberg exchange coupling is the dominant spin-spin interaction. In addition to the exchange interaction, there is single-ion anisotropy with the direction of the easy axis being the same for all spins. The analysis is carried out in the limit $S >> 1$, where $S$ denotes the ionic spin, and in the temperature range $T << J_{ave}S$, where $J_{ave}$ is the average magnitude of the exchange interaction. It is assumed that the independent boson picture is appropriate so that the thermodynamic properties of the chain are those of a gas of weakly interacting magnons whose frequencies are obtained from linearized equations of motion for the spins. The $\pm J$ model is studied in detail with both random and non-random anisotropy. Particular emphasis is placed on the existence and magnitude of the anisotropy gap and its effect on the specific heat and the magnetic susceptibility.

## I. Introduction

The low-temperature properties of disordered magnetic systems has been a subject of interest for some time. In a series of recent publications, particular attention was paid to the dynamical properties of one-dimensional disordered magnets [1 - 9]. The analyses reported in these references were based on linearizing the equations of motion of the spins about a classical ground state. The eigenfrequencies of the equations of motion were identified as magnon energies ($\hbar = 1$) so that the resulting thermodynamic properties were those of an ideal boson gas. In the boson picture, the free energy and thus all of the thermodynamic properties can be expressed as integrals over the magnon density of states. The density of states was obtained by mode counting techniques and transfer-matrix scaling arguments and compared with the predictions of approximate analytical calculations based on the coherent exchange approximation. For quantum spin systems, the description in terms of linearized excitations is generally valid only in the limit $S \gg 1$. Complementary to this work are theoretical studies of disordered quantum spin chains with $S = 1/2$ utilizing various approaches such as series expansion, transfer matrix and real space renormalization group techniques [10 - 13]. In Refs.1 - 9, work was carried out on a variety of systems including the Heisenberg and XY models in zero and finite applied fields, while in Refs. 10 - 13, the analysis of the quantum systems was limited to the spin-1/2 and spin-1(Ref. 11) Heisenberg models in zero field. Also relevant are the studies of disordered *classical* spin chains with the $\pm J$ model Hamiltonian that are reported in Ref. 10. However, none of the references cited have addressed the thermodynamic properties of weakly *anisotropic* systems, where there is a small anisotropy term in the Hamiltonian in addition to the dominant isotropic Heisenberg interaction. In this paper, we will investigate the effects of anisotropy in the large-*S* limit for a system with the Hamiltonian

$$H = -\sum_n J_{n,n+1} \vec{S}_n \cdot \vec{S}_{n+1} - \sum_n D_n S_{nz}^2 - H \sum_n S_{nz} \qquad (1)$$

Here, in addition to the Heisenberg interaction, there is an anisotropic single-ion term and a uniform applied field. We consider only the case of uniaxial anisotropy with the anisotropy (z) axis the same for all of the ions, which are assumed to have the same value of *S*. In the most general case, both the exchange integrals and the anisotropy constants are random functions of position. We assume that the mean values obey the inequality <|*J*|> >> <*D*> > 0 consistent with the picture of weak, easy-axis anisotropy. The applied field is parallel to the anisotropy axis and satisfies the condition *H* << <*D*>.

In three dimensional ferromagnetic and antiferromagnetic systems with long range order, easy axis anisotropy produces a gap in the density of states of the magnons that has a pronounced effect on the behavior of the specific heat and the susceptibility at low temperatures. In this paper, we address the question of whether a similar effect occurs in one-dimensional disordered systems. As a specific example, we study the ±*J* model, where the sign of the exchange interaction is a random variable, as is the value of the anisotropy constant. Our approach is to use negative eigenvalue counting techniques [3,4,6-8] to determine the magnon density of states in the low energy region *E* << <|*J*|>S. From the behavior of the density of states, we infer the limiting behavior of the specific heat and the longitudinal susceptibility for the case where the uniform field is along the anisotropy axis.

The rest of the paper is organized as follows. The linearized equations of motion for the transverse components of the spin are developed in Sec. II. In Sec. III we report detailed results for the densities of states of the anisotropic ±$J$ model and comment on their effect on the thermodynamic properties. The broader implications of our findings are discussed in Sec. IV.

## II. Equations of motion and thermodynamic functions

With the Hamiltonian displayed in Eq. (1), the Fourier transforms of the linearized equations of motion for the transverse spin operators, $S_{n+}$, take the form

$$(\omega - H - 2D_n <S_{nz}>_0 - J_{n,n+1} <S_{n+1z}>_0 - J_{n-1,n} <S_{n-1z}>_0)S_{n+}$$

$$= -J_{n,n+1} <S_{nz}> S_{n+1+} - J_{n-1,n} <S_{nz}> S_{n-1+} \quad (2)$$

Here the angular brackets $<\ldots>_0$ refer to the expectation values of the z component of the spin in the classical (Néel) ground state where the orientation of the spin is determined by the sign of the nearest-neighbor interaction:

$$<S_{nz}>_0 = \text{sign}(J_{n,n-1}) <S_{n-1z}>_0. \quad (3)$$

Broadly speaking, we can divide the ground states into two categories: if there are equal numbers of 'up' and 'down' spins in the presence of an infinitesimal applied field, we designate the system as 'antiferromagnetic', whereas if there is a macroscopic ($O(N)$) difference in the numbers of up and down spins, we refer to the system as 'ferrimagnetic' for which complete (or ferromagnetic) alignment is a limiting case.

For future reference, we note that when there is no disorder and a ferromagnetic interaction between the spins ($J > 0$), the characteristic frequencies, or magnon energies, have the simple form

$$\omega_k = H + 2DS + 2JS[1 - \cos(k)], \quad -\pi \leq k \leq \pi \quad (4)$$

In the case where there is an antiferromagnetic interaction between the spins ($J < 0$), the characteristic frequencies come in pairs:

$$H \pm 2S\left[(J+D)^2 - J^2 \cos^2(k/2)\right]^{1/2}$$

In the standard interpretation, the states with negative frequency are identified with magnon modes having energies equal to the magnitude of the frequency. As a result, one has two bands of magnons with energies differing by 2H:

$$\omega_k^\pm = 2S\left[(J+D)^2 - J^2 \cos^2(k/2)\right]^{1/2} \pm H, \quad -\pi \leq k \leq \pi \quad (5)$$

In the presence of disorder, the solution to the equations of motion will yield both positive and negative frequencies (except in the limit of complete ferromagnetic alignment). As in the case of the ideal antiferromagnet, we identify the negative frequencies with magnon modes with energies equal to the magnitude of the characteristic frequencies so that we can write

$$\omega_j^+(H) = \omega_{j0}^+ + H \quad (6)$$

$$\omega_j^-(H) = \omega_{j0}^- - H \quad (7)$$

where the $\omega_{j0}^\pm$ denote the magnon energies in the absence of an applied field. Note that it is assumed that the $H$ is smaller than any of the $\omega_{j0}^-$ so that all of magnon energies are positive, as required for stability.

In the consideration of the thermodynamic functions, it is useful to introduce the densities of states associated with the $\omega_{j0}^+$ and $\omega_{j0}^-$, which we denote by $\rho^+(\omega)$ and $\rho^-(\omega)$, respectively. In the zero-field limit, the specific heat has the form

$$C(T) = T^{-2} \int_0^\infty \omega^2 [\rho^+(\omega) + \rho^-(\omega)] e^{\omega/T} (e^{\omega/T} - 1)^{-2} d\omega \qquad (8)$$

The analysis of the magnetic properties depends on whether the system is ferrimagnetic or antiferromagnetic. In the former case, we assume an infinitesitmal applied field and consider the change in the magnetic moment with temperature, denoted by $\Delta M(T)$. In the limit as $H \to 0$, we obtain

$$\Delta M(T) = -\int_0^\infty [\rho^+(\omega) - \rho^-(\omega)](e^{\omega/T} - 1)^{-1} d\omega \qquad (9)$$

For an antiferromagnet, one has $\rho^+(\omega) = \rho^-(\omega)$ with the result that $\Delta M(T) = 0$, and there is no magnetization in an infinitesimal field. However, the longitudinal zero-field susceptibility, $\chi(T)$, is finite and is given by the expression

$$\chi(T) = T^{-1} \int_0^\infty [\rho^+(\omega) + \rho^-(\omega)] e^{\omega/T} (e^{\omega/T} - 1)^{-2} d\omega \qquad (10)$$

Equations (7 – 9) show that at low temperatures, the specific heat, the change in magnetization and the susceptibility depend critically on the behavior of the density of states in zero applied field as $\omega \to 0$. If there is a gap, as happens in the absence of disorder when $D \neq 0$, all of these functions will vary exponentially as $T \to 0$. The question then is what is the effect of disorder on the density of states and is there a gap in the presence of disorder. We will take this up in the next section using the $\pm J$ model to simulate the effects of disorder in an antiferromagnetic system.

## III. ±*J* Model

In this section we present results for the ±*J* model with random anisotropy. The sign of the exchange interaction has the distribution

$$P[sign(J_{n+1,n})] = (1-c)\delta[sign(J_{n+1,n}) - 1] + c\delta[sign(J_{n+1,n}) + 1] \qquad (11)$$

Thus when $c = 0$, all of the interactions are ferromagnetic and when $c = 1$, the array is a perfect antiferromagnet. For $0 < c < 1$ one also has an antiferromagnet with properties such that when $c \approx 0$ there are large regions of parallel spins, and when $c \approx 1$ there are large regions of alternating antiparallel spins. We also take into account the effects of disorder in the anisotropy parameters by assuming that $D_n$ is uniformly distributed between 0 and 0.01*J*.

The density of states of the ±*J* model in the absence of anisotropy has been investigated previously with exact and numerical results obtained for $c = ½$ [3], and numerical results and an accurate analytical approximation for $c \neq ½$ [8]. For all values of $c \neq 0,1$, the density of states varied as $\omega^{-1/3}$ in the limit $\omega \to 0$. The –1/3 power law contrasts with constant density of states for the isotropic antiferromagnet and the $\omega^{-1/2}$ variation characteristic of the isotropic ferromagnetic chain.

We have used negative eigenvalue counting techniques [3,8] to determine the *integrated density of states* (IDOS) for chains of $10^7$ spins with $c = 0.25, 0.50,$ and $0.75$. The only change required in the formalism is that Eq. (1.6) of Ref. 8 is modified to read

$$(2 + (2D_j / J) - z_j \omega)V_j = V_{j+1} + V_{j-1}$$

where the symbols $z_j$ and $V_j$ are defined in that reference. The results for the IDOS are shown in Fig. 1. Note that the IDOS($\omega$) is the number of modes per spin with frequencies in the interval between 0 and $\omega$ and thus corresponds to $N^{-1}\int_0^\omega \rho^+(\omega')d\omega'$ when $\omega > 0$ and to $N^{-1}\int_\omega^0 \rho^-(\omega')d\omega'$ when $\omega < 0$, so that in general, one has IDOS($\infty$) + IDOS($-\infty$) = 1.

It is evident from the figure that the integrated density of states is symmetric, as expected for an antiferromagnet, and shows a gap that depends on $c$. Because of the presence of a few modes with energies very close to zero (*i.e.* band tailing), it is impossible to give a precise definition of the gap. As a consequence, we take an 'operational' point of view and define the band gap as the energy (in units of $JS$) where IDOS = 0.0005. Using this definition, our results for the gap as a function of c are shown in Fig. 2. Also shown are the limiting values of the gap for the ferromagnetic chain ($c = 0$) and the antiferromagnetic chain ($c = 1$). Here the effect of the random anisotropy is to reduce the values of the gaps relative to those of the ideal ferromagnetic and antiferromagnetic chains with the common, site-independent value $D = <D> = 0.005J$, where the gaps are equal to $2DS = 0.01JS$ and $2^{3/2}S(JD)^{1/2} = 0.20JS$ ($D \ll J$), respectively. With a distribution of anisotropy parameters, the gaps are .077$JS$ ($c = 0$) and 0.17$JS$ ($c = 1$). In the intermediate case where $c = ½$, the gap with $D = 0.005J$ is 0.017$JS$, whereas with random $D$ it falls to 0.015$JS$. Thus the disorder in $D$ reduces the gap by a factor ~ 0.8 – 0.9.

We have also studied the variation of the gap with the strength of the anisotropy. With a common value for the anisotropy constant, we find that the gap varies as $D^{3/4}$. Such behavior is to be expected from scaling arguments that exploit the similarity with the XY ferromagnetic chain in a random transverse field [2,9]. The $D^{3/4}$ variation falls between the linear variation of the gap in ideal ferromagnetic chains and the square root behavior for the ideal antiferromagnetic chain ($D \ll J$).

In the analysis of the behavior of the specific heat and the susceptibility of weakly anisotropic magnets at low temperatures, there are two limiting regimes to consider: $\Delta \ll T \ll JS$ and $T \ll \Delta$, where $\Delta$ denotes the anisotropy gap. When $T \ll \Delta$, both the susceptibility and the specific heat vary as $\exp[-\Delta/T]$. In the range $\Delta \ll T \ll JS$, the gap has a negligible effect on the specific heat and one has $C(T) \sim T^{2/3}$ [7]. In contrast, even when $\Delta \ll T \ll JS$, the susceptibility integral, Eq. 8, is strongly influenced by the gap. One finds that $\chi$ is a linear function of the temperature, *i.e.*

$$\chi(T) \approx T \int_0^\infty [\rho^+(\omega) + \rho^-(\omega)] \omega^{-2} d\omega \qquad (12)$$

## IV. Discussion

The principal result emerging from this work is that there can be an easy-axis anisotropy gap even in strongly disordered magnetic chains. Although the disorder renormalizes the gap (*cf*. Fig. 2) and introduces mid-gap states, it does not do away with the gap altogether. The effect of introducing 'wrong sign' exchange interactions in an otherwise perfect array is to shift the gap towards the value it would have if all of the interactions were of the opposite sign. At least for the model we studied, the effect of disorder in the strength of the anisotropy is rather small (gap renormalization ~ 0.8 – 0.9). It is also important to note that our analysis pertains only to situations where the $D_n$ are greater or equal to zero. Negative values of $D_n$, which correspond to local easy-plane anisotropy, give rise to modes with imaginary frequencies, indicating an instability in our hypothesized easy-axis ground state.

The effect of the anisotropy gap on the thermodynamic properties is pronounced at low temperatures where it gives rise to exponential behavior. At temperatures well above the gap, the specific heat is not significantly affected by the anisotropy. This does not happen, however, in the case of the susceptibility. Here the anisotropy is needed to stabilize the array in the presence of an applied field. Without anisotropy, the susceptibility diverges in the large-$S$ limit.

The divergence of the susceptibility of the isotropic system is a shortcoming of the ideal boson approximation. This divergence reflects the relatively large effect of the applied field on the low frequency modes which are all shifted by an amount $\Delta\omega = H$ independent of the value of $\omega$. What is missing in the approximation is the influence of the thermal fluctuations, which are strong in one dimension. The role of the fluctuations is to reduce the long-range coherence between the spins that is implicit in the ideal boson approximation for the low frequency modes. The destruction of the coherence gives rise to the Curie-like ($\chi \propto T^{-1}$) behavior of the susceptibility in the limit $T \to 0$ that was found for quantum and classical disordered chains in the analyses reported in Refs. 10-13. It should be noted that objections to the use of the boson approximation for isotropic systems appear to apply primarily to the calculation of the longitudinal susceptibility. The results for the specific heat are expected to be more reliable since the contribution from the very low frequency modes is suppressed by the factor of $\omega^2$ appearing in the integrand in Eq. (8). When these modes are suppressed, the integral converges and one obtains power law behavior for the specific heat similar to that found in Ref. 12 [8].

Concerning the anisotropic systems, it is our expectation that the anisotropy reduces the role of the fluctuations so that the analysis of the susceptibility in the ideal boson approximation is more realistic. To test such a hypothesis one would need to apply real space renormalization group and related techniques to anisotropic disordered chains. It is particularly important to determine whether the susceptibility and the specific heat vary as $\exp[-\Delta/T]$ when $T << \Delta$. When $\Delta << T << JS$, we predict a linear temperature dependence for the susceptibility and a $T^{3/2}$ variation for the specific heat. As noted, the power law behavior of the specific heat that is characteristic of the isotropic system in the large-$S$ limit is also found for the isotropic spin-1/2 random chain [12], where the corresponding exponent $\approx 0.44$. Whether the susceptibility of the weakly anisotropic spin-1/2 random chain varies linearly with T over an appreciable part of the range $\Delta << T << JS$ is yet to be determined.

Finally we note that while our conclusions were drawn from a study of a somewhat artificial model, the linearized equation of motion together with eigenvalue counting techniques can be used to study the densities of states in more realistic models formulated for materials of interest.

**Acknowledgment** This work is partially sponsored by the Scientific and Technical Council of Turkey (TUBITAK).

## Figure Captions

Fig. 1. Integrated density of states (IDOS) versus energy for the $\pm J$ model
From top to bottom, the curves correspond to $c = 0.25$, $0.50$ and $0.75$.
Energy is in units of $JS$, and $D$ is uniformly distributed between 0
and $0.01J$.

Fig. 2. Gap versus fraction of antiferromagnetic interactions. The gap is
in units of $JS$, and $D$ is uniformly distributed between 0 and $0.01J$.
Results obtained with a fixed value of $D \equiv \langle D \rangle = 0.005J$ are larger by a
factor of $1.15 - 1.30$.

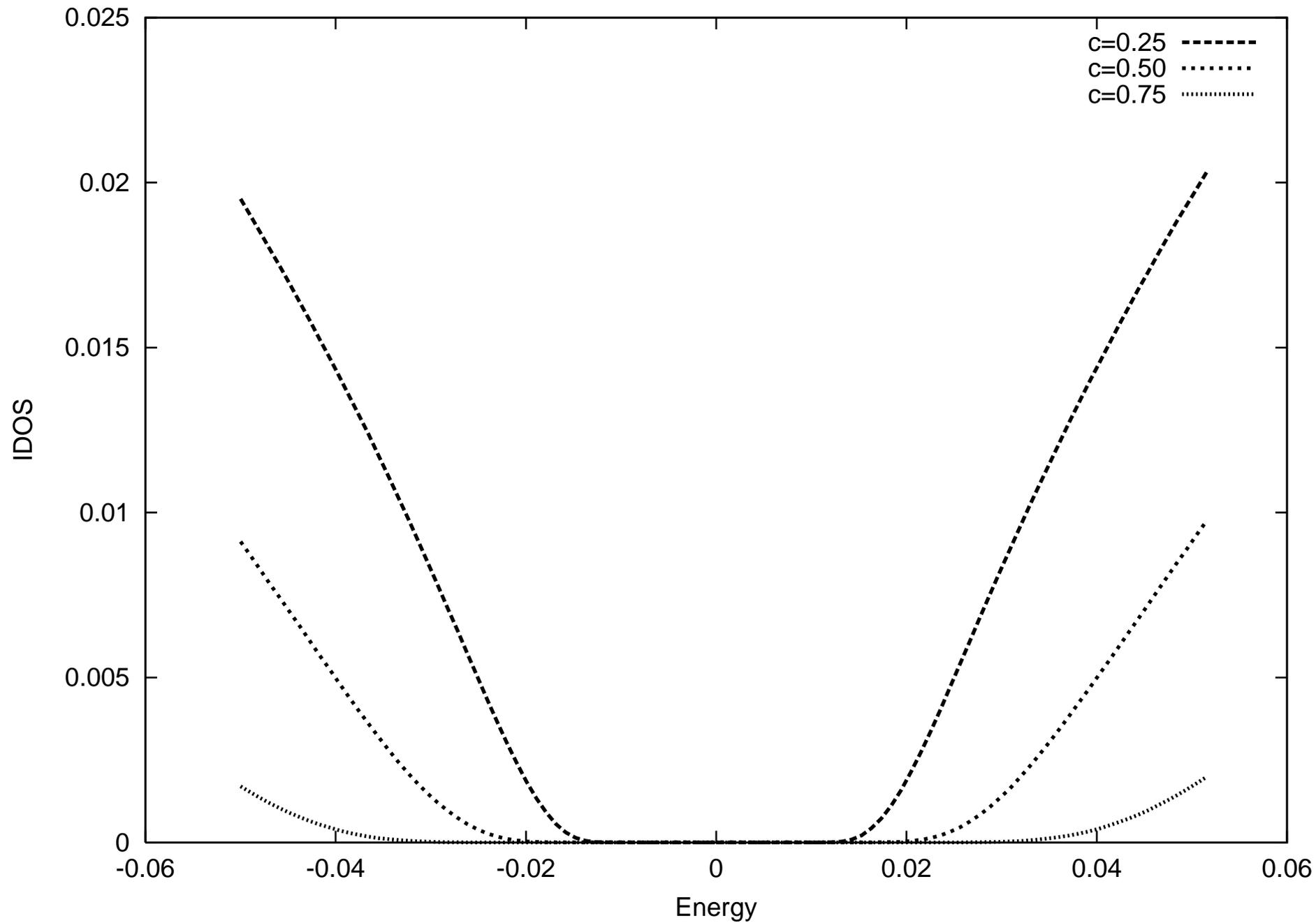

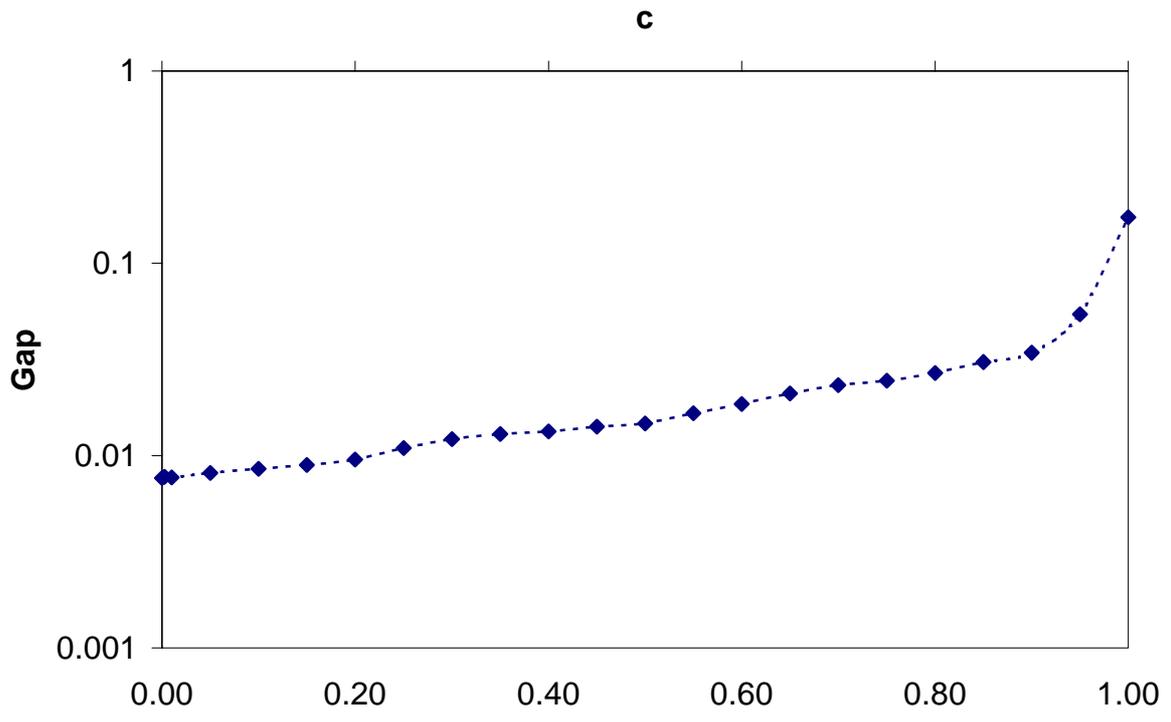